\documentclass[12pt,a4paper]{iopart} 
\usepackage{graphicx,cite}
\eqnobysec
\begin{document}

\title[Bose-Einstein condensation dynamics]{Bose-Einstein condensation
dynamics from the numerical solution of the Gross-Pitaevskii equation}

\author{Sadhan K. Adhikari and Paulsamy Muruganandam}

\address{Instituto de F\'{\i}sica Te\'orica, Universidade Estadual Paulista,
01.405-900 S\~ao Paulo, S\~ao Paulo, Brazil}

\begin{abstract}

We study certain stationary and time-evolution problems of trapped
Bose-Einstein condensates using the numerical solution of the Gross-Pitaevskii
equation with both spherical and axial symmetries. We consider time-evolution
problems initiated by changing the interatomic scattering length or harmonic
trapping potential suddenly in a stationary condensate. These changes introduce
oscillations in the condensate which are studied in detail. We use a time
iterative split-step method for the solution of the time-dependent
Gross-Pitaevskii equation, where all nonlinear and linear nonderivative terms
are treated separately from the time propagation with the kinetic energy
terms. Even for an arbitrarily strong nonlinear term this leads to extremely
accurate and stable results after millions of time iterations of the original
equation.

\end{abstract}


\submitto{\jpb}

\maketitle

\section{Introduction}
 
Since the successful detection \cite{1} of Bose-Einstein condensates 
in dilute weakly-interacting bosonic atoms employing magnetic trap at
ultra-low temperature, there have been intense theoretical studies on
different aspects of the condensate
\cite{11,s1,s21,s2,s5,s31,s32,s3,s4,2,3x,3y}. The
experimental
magnetic trap could be either spherically symmetric or axially symmetric.  
The properties of an ideal condensate at zero temperature are usually
described by the time-dependent, nonlinear, mean-field Gross-Pitaevskii
(GP) equation \cite{8} which incorporates appropriately the trap potential
as well as the interaction between the atoms forming the condensate.
There are many numerical methods for the solution of the GP equation
\cite{s1,s21,s2,s5,s31,s32,s3,s4}.
The effect of the interatomic interaction leads to the nonlinear term in
the GP equation which complicates the solution procedure. Also, to
simulate the proper experimental situation one should be prepared to deal
with an axially-symmetric harmonic oscillator trap in addition to a
spherically symmetric one \cite{s31,s32,s3,s4}.

A numerical study of the time-dependent GP equation is extremely
interesting from a physical point of view, as this can provide solution to
many stationary as well as time-evolution problems involving the
condensate. The stationary problems are governed by a wave function with
trivial time ($\tau$) dependence $\Psi ({\bf r}, \tau) = \Psi ({\bf
r}) \exp (-i\mu \tau/\hbar)$, so that $|\Psi ({\bf r}, \tau)| = |\Psi
({\bf
r})|,$  where $\mu $  is a real energy parameter. Such problems with
trivial time dependence and under the action of a spherically symmetric
trap  can 
be treated by time-independent Runge-Kutta integration method
\cite{s1}.  
The solution of these problems can also be extracted from the solution of
the time-dependent GP equation \cite{s21,s2}. In the presence of an
axially-symmetric trap, 
there is a  time-independent expansion scheme for
the solution of the stationary problem \cite{s5}. However, for most
applications  one solves
the
time-dependent GP equation and extract the stationary wave function 
$\Psi ({\bf r})$ and the parametric energy $\mu$ \cite{s3,s4}. 
In addition to obtaining the solution of the stationary
problem, the time-dependent GP equation can be used to
study the intrinsic time-evolution problems with nontrivial time
dependence.

There are several methods for the numerical solution of the time-dependent
GP equation. Most  of them  adopt a similar time-iteration procedure
in execution,
although the details could be different. The time-dependent
GP equation is first discretized in space and time and then solved
iteratively with an initial input solution \cite{s21,s2,s3,s4}.
In the spherically symmetric case the GP equation is solved
by discretization with  the Crank-Nicholson scheme \cite{ames,dl} for time 
propagation
complimented by the
known
boundary conditions \cite{s2}. However, in the axially symmetric case 
a two-step procedure \cite{ames} is first used to separate the radial and
axial parts
of the Hamiltonian before applying the   the explicit \cite{s31}  or
the Crank-Nicholson scheme \cite{s32,s3,s4} for time propagation in
each
step. In all these approaches the nonlinear term is treated together with  
the
explicit or
the
Crank-Nicholson scheme \cite{s21,s2,s32,s3,s4}.
It is well known that such a Crank-Nicholson method may suffer from a
possible numerical amplification of random numerical noise, due to
nonlinearity, leading to spurious input of kinetic energy and limiting the
stability of the algorithm to short time (small number of
iterations) \cite{s32}. This also limits the applicability of the method
only to
small
values of the nonlinearity \cite{s21}.

In this paper we suggest a different split-step time-iteration method for the
solution of the time-dependent GP equation where the full Hamiltonian is split
into three parts \cite{ames}. The nonlinear  as well as the different linear
(nonderivative) terms (excluding spatial derivatives)   are treated in one
step. In the spherical case the spatial derivative is treated in another  step.
In the axially symmetric case the radial and axial derivatives are dealt with
in two separate steps.  In the treatment of nonderivative terms  the numerical
error is  kept to a minimum. In dealing with the derivative kinetic energy
terms  the GP equation is solved by discretization using the Crank-Nicholson
scheme for time propagation  complimented by the known boundary conditions
\cite{s4,dl,koo}. The advantage of the present split-step procedure  is that
the nonlinear and other linear nonderivative terms can be  treated very
precisely  and this improves the accuracy and the stability of the method
compared to previous ones \cite{s4}. Consequently, we could easily find
accurate solution of the GP equation for very large nonlinearity, which 
remains stable even after millions of iterations. This will be of advantage in
using the present approach for the study of time-evolution problems during a
large interval of time.

As an application of the present approach in studying time-evolution problems
we consider in the spherically symmetric case the following two types of
problems: (1) On a previously formed condensate the harmonic oscillator
trapping potential  is increased or decreased suddenly by a factor of
two. This
can be achieved in the laboratory by changing the current in the coil
responsible for the magnetic field responsible for trapping.  (2)  On a
previously formed condensate the atomic scattering length is increased
or decreased suddenly by a factor of two. Case (2) above is of interest as
recently it
has been possible to modify the scattering length in a controlled fashion by
exploiting a Feshbach resonance \cite{fb1,fb2}.  In both cases we study the
resultant oscillation of the root mean square radius  of the condensate and its
first time derivative. In the axially symmetric case we also  consider the
time-evolution problems above and study the resultant oscillation  of the root
mean square sizes in radial and axial directions. 

In section 2 we describe briefly the time-dependent  GP equation with 
spherical and axial traps. The split-step Crank-Nicholson method for time
propagation is described in section 3.  In sections 4 and 5 we report the
numerical results of the present investigation for the spherically and axially
symmetric cases, respectively. Finally, in sections 6 and 7 we present 
a discussion and summary of our study.

\section{Nonlinear Gross-Pitaevskii Equation}

At zero temperature, the time-dependent Bose-Einstein condensate wave
function $\Psi({\bf r};\tau)$ at position ${\bf r}$ and time $\tau $ may
be described by the following  mean-field nonlinear GP equation
\cite{11,8}
\begin{eqnarray}\label{a} \biggr[ -\frac{\hbar^2\nabla ^2}{2m}
&+& V({\bf r})  
+ gN|\Psi({\bf
r};\tau)|^2
-i\hbar\frac{\partial
}{\partial \tau} \biggr]\Psi({\bf r};\tau)=0.   \end{eqnarray} Here $m$
is
the mass and  $N$ the number of atoms in the
condensate, 
 $g=4\pi \hbar^2 a/m $ the strength of interatomic interaction, with
$a$ the atomic scattering length. The normalization condition of the wave
function is
$ \int d{\bf r} |\Psi({\bf r};\tau)|^2 = 1. $

\subsection{Spherically Symmetric Case}

In this case  the trap potential is given by $  V({\bf
r}) =\frac{1}{2}m \omega ^2r^2$, where $\omega$ is the angular frequency
and $r$ the radial distance. The wave function can be written as
$\Psi({\bf r};\tau)=\psi(r,\tau)$. 
 After a transformation of variables to 
dimensionless quantities 
defined by $x =\sqrt 2 r/l$,     $t=\tau \omega, $
$l\equiv \sqrt {(\hbar/m\omega)} $ and
$\phi(x,t)\equiv \varphi(x,t)/x =\psi(r,\tau)(4 
\pi l^3)^{1/2}$, the GP equation in this case becomes
\begin{eqnarray} \label{c}
\left[-\frac{\partial^2}
{\partial x^2}+\frac{x^2}{4}+{\cal N}\left|
\frac{\varphi(x,t)}{x}
\right| ^2 -i\frac{\partial }{\partial t}\right] \varphi (x,t)=0, 
\end{eqnarray} 
where ${\cal N}=Na/l$. The normalization condition for the wave function is 
\begin{eqnarray}\label{n1}
\int_0^\infty dx |\varphi(x,t)|^2=2\sqrt 2.
\end{eqnarray}

\subsection{Axially Symmetric Case}

The trap potential is given by  $  V({\bf r}) =\frac{1}{2}m \omega
^2(\rho^2+\lambda^2 z^2)$  where  $\omega$ is the angular frequency in the
radial direction $\rho$ and  $\lambda \omega$ that in  the axial direction $z$.
We are using the cylindrical coordinate system ${\bf r}\equiv (\rho,\theta,z)$
with $\theta$ the azimuthal  angle. In this case one can have quantized vortex
states with rotational motion around the $z$ axis. In such a vortex the atoms
flow with tangential velocity $\hbar L/(mr)$ such that each atom has quantized
angular momentum $\hbar L$ along $z$ axis.  The wave function can then be
written as: $ \Psi({\bf r};\tau)=\psi(r,z;\tau)\exp (iL\theta),$ with $L=0, \pm
1, \pm 2, ...$.
 
Using the above angular distribution of the wave function in (\ref{a}),  in
terms of dimensionless variables $x =\sqrt 2 r/l$,  $y=\sqrt 2 z/l$,   $t=\tau
\omega, $  $l\equiv \sqrt {\hbar/(m\omega)}$,  and   $ \phi(x,y;t)\equiv  {
\varphi(x,y;t)}/{x} =  \sqrt{4\pi l^3}\psi(r,z;\tau), $ we get
\begin{eqnarray}\label{d}
\fl \biggr[ -\frac{\partial^2}{\partial
x^2} + \frac{1}{x}\frac{\partial}{\partial x} -\frac{\partial^2}{\partial
y^2}+\frac{L^2}{x^2}
+\frac{1}{4}\left(x^2+\lambda^2 y^2-\frac{4}{x^2}\right) 
& + & {\cal N}\left|\frac {\varphi({x,y};t)}{x}\right|^2 \nonumber \\
& & -i\frac{\partial }{\partial t} \biggr]\varphi({ x,y};t) = 0.
\end{eqnarray}
The normalization condition  of the wave
function is \begin{equation}\label{5}  \int_0 ^\infty
dx \int _{0}^\infty dy|\varphi(x,y;t)|
^2 x^{-1}=2\sqrt 2. \label{n2} \end{equation}

\section{Split-Step Crank-Nicholson Method}

The GP equation in both spherically symmetric and axially symmetric cases
can be formally written as 
\begin{equation}\label{gp1}
i \frac{\partial \varphi}{\partial t} = H \varphi,
\end{equation}
where the Hamiltonian $H$ contains the different nonlinear and linear
terms including the spatial derivatives. We solve this equation by
iteration \cite{dl,ames,koo}. A given  trial input solution is propagated
in time
over small
time steps until a stable final  solution is reached. The GP equation is
discretized in space and time using the finite difference scheme. This
procedure results in a set of algebraic equation which can be solved by
time
iteration using an input solution consistent with the known boundary
condition. In the present split-step method \cite{ames} 
this iteration
is conveniently done in several steps by breaking up the full
Hamiltonian into different derivative and non-derivative parts.

In the
spherically symmetric case we split $H$ in to three
parts: $H=H_1+H_2+H_3,$
where 
\begin{eqnarray}
H_1&=& \frac{1}{2} \left[ \frac{x^2}{4}+{\cal N}\left|
\frac{\varphi(x,t)}{x}
\right| ^2 \right], \\ \label{h2}
H_2&=&-\frac {\partial ^2}{\partial x^2}, \\
H_3 &=& H_1. 
\end{eqnarray}
The time variable is discretized as $t_n=n\Delta$ where $\Delta$ is  the time
step. The solution is advanced first over the time step  $\Delta$ at time $t_n$
by solving the GP equation (\ref{gp1}) with $H=H_1$  to produce an intermediate
solution $\varphi^{n+1/3}$ from $\varphi^n$, where $\varphi^n$ is the
discretized wave  function at time $t_n$. As there is no derivative in $H_1$ 
this propagation is performed essentially exactly  for small $\Delta$ through
the operation
\begin{eqnarray}\label{al1}
\varphi^{n+1/3}
&=& {\cal O}_{\mbox{nd}}(H_1) \varphi^n \equiv  e^{-i\Delta H_1}
\varphi^n,
\end{eqnarray}
where ${\cal O}_{\mbox{nd}} (H_1)$ denotes time-evolution operation with $H_1$
and the suffix `nd' denotes non-derivative. Next we perform the time
propagation corresponding to the operator $H_2$ numerically by the following
semi-implicit Crank-Nicholson scheme \cite{dl}:
\begin{equation}\label{gp2}
\frac{ \varphi^{n+2/3}- \varphi^{n+1/3}}{-i\Delta } = \frac{1}{2}H_2(
\varphi^{n+2/3}+ \varphi^{n+1/3}).\label{a1}
\end{equation}
The formal solution to (\ref{gp2}) is 
\begin{equation}\label{gp3}
 \varphi^{n+2/3}= {\cal O}_{\mbox{CN}}(H_2) \varphi^{n+1/3}
\equiv 
\frac{1-i\Delta H_2/2          }{ 1+i\Delta
H_2/2 }
\varphi^{n+1/3},
\end{equation}
where ${\cal O}_{\mbox{CN}} $ denotes time-evolution operation 
with $H_2$ and   the
suffix `CN' refers to the  Crank-Nicholson algorithm.
Operation  ${\cal O}_{\mbox{CN}} $ 
is used to propagate the intermediate solution  $  \varphi^{n+1/3}
$ by time step $\Delta$  to generate the second intermediate  solution 
$\varphi^{n+2/3}$. The final solution 
is  obtained from  
\begin{equation}\label{gp4}
 \varphi^{n+1}=  {\cal O}_{\mbox{nd}}(H_3) {\cal O}_{\mbox{CN}}(H_2){\cal
O}_{\mbox{nd}}(H_1) 
\varphi^{n}.
\end{equation}
The break-up of the nonderivative term in two parts $- H_1$ and $H_3 -$ 
symmetrically around
the derivative term $H_2$, increases enormously the stability of the
method and reduces the numerical error.

In the axially symmetric case $H$ is split into three parts: $H
=H_1+H_2+H_3$, where 
\begin{eqnarray}
H_1&=&\frac{1}{4}\left(x^2+\lambda^2 y^2-\frac{4}{x^2}\right)
+\frac{L^2}{x^2}+{\cal N}\left|
\frac{\varphi(x,y,t)}{x}
\right| ^2 ,
\\ \label{h22}
H_2&=&-\frac {\partial ^2}{\partial x^2}+\frac{1}{x}\frac{\partial}
{\partial x},
\\ \label{h3}
H_3&=&-\frac {\partial ^2}{\partial y^2}.
\end{eqnarray}
The strategy is now obvious. 
The solution is advanced first over the time step 
$\Delta$ at time $t_n$ using the Hamiltonian $H_1$ above to produce an
intermediate solution $\varphi^{n+1/3}$ from $\varphi^n$. 
For small $\Delta$ this propagation is performed
essentially exactly  via (\ref{al1}).
Next we perform two successive  time propagations corresponding to the
operators 
$H_2$ and $H_3$ numerically by the Crank-Nicholson scheme (\ref{gp3}).
Consequently,  a single time iteration from $t_n$  to $t_{n+1}$ is
performed via the following three-step operation:
\begin{equation}
 \varphi^{n+1}=  {\cal O}_{\mbox{CN}}(H_3) {\cal O}_{\mbox{CN}}(H_2){\cal
O}_{\mbox{nd}}(H_1) \varphi^n.
\end{equation}
In the axially symmetric case due to the intrinsic asymmetry of the
kinetic energy terms in $x$ and $y$ directions, the complete
symmetrization as in the spherically symmetric case (\ref{gp4}) is
nontrivial and we
did not attempt such a symmetrization. 

The advantage of the above split-step method with small time step $\Delta$
is due to the following three factors \cite{ames,dl}. First, all
iterations
conserve normalization of the wave function. Second, the error involved in
splitting the Hamiltonian is proportional to $\Delta^2$ and can be
neglected and the method preserves the symplectic structure of the
Hamiltonian formulation.  Finally, as a major part of the Hamiltonian
including the nonlinear term is treated fairly accurately without mixing
with the Crank-Nicholson propagation, the method can deal with an
arbitrarily large nonlinear term and lead to stable and accurate converged
result.

Now we describe the Crank-Nicholson algorithm in the
spherically- and axially-symmetric cases.  
In the spherically symmetric case the GP equation is mapped on to  $N_x$
one-dimensional grid points in $x$. Equation (\ref{gp1}) is discretized
with $H=H_2$ of (\ref{h2}) 
by the following Crank-Nicholson scheme \cite{ames,dl,koo}:
\begin{eqnarray}\label{kn1}
\fl \frac{i(\varphi_{j}^{n+1}-
\varphi_{j}^n)}{\Delta}=-\frac{1}{2h^2}\biggr[(\varphi^{n+1}_{j+1}-
2\varphi^{n+1}_j
+\varphi^{n+1}_{j-1})
+(\varphi^{n}_{j+1}-2\varphi_{j}^{n}
+\varphi^{n}_{j-1})\biggr],
\end{eqnarray}
where $\varphi_j^n=\varphi(x_j,t_n)$ refers to  $x=x_j=jh,$
$j=1,2,...,N_x$ and
$h$ is the space step. This scheme is constructed by approximating
$\partial
/\partial t$ by a two-point formula and $\partial^2
/\partial x^2$ by a three-point formula. 
This
procedure results in a series of tridiagonal sets of equations
(\ref{kn1}) in $\varphi^{n+1}_{j+1}$,  $\varphi^{n+1}_{j}$, and
$\varphi^{n+1}_{j-1}$ at time $t_{n+1}$,
 which are solved using the proper boundary conditions.

The Crank-Nicholson scheme (\ref{kn1}) possesses certain properties worth
mentioning \cite{dl,ames}. This error in this scheme is both second order
in
space and time steps 
so that for
small $\Delta$ and $h$ the error is negligible. This scheme is also
unconditionally stable. The boundary condition at infinity is preserved 
for small values of $\Delta/h^2$\cite{dl}.

In the axially symmetric case the time-dependent GP equation 
(\ref{d}) is
mapped on a two-dimensional grid of points $N_x\times N_y$ in $x$ and $y$.
Equation (\ref{gp1}) with $H=H_2$ of (\ref{h22}) is discretized using
the following  Crank-Nicholson scheme
\cite{ames,dl,koo}: 
\begin{eqnarray}\label{kn2}
\fl \frac{i(\varphi_{j,p}^{n+1}-
\varphi_{j,p}^n)}{\Delta}
& = & -\frac{1}{2h^2}\biggr[(\varphi^{n+1}_{j+1,p}-2\varphi_{j,p}^{n+1}
+\varphi^{n+1}_{j-1,p}) 
+(\varphi^{n}_{j+1,p}-2\varphi_{j,p}^{n}
+\varphi^{n}_{j-1,p})\biggr]\nonumber \\ & & 
+\frac{1}{4x_jh}\left[(\varphi^{n+1}_{j+1,p}-\varphi^{n+1}_{j-1,p})
+(\varphi^{n}_{j+1,p}-\varphi^{n}_{j-1,p})  \right], \end{eqnarray} 
where the discretized wave function $\varphi^n_{j,p}\equiv 
\varphi(x_j,y_p,t_n)$ refers to a fixed $y=y_p=ph$, $p=1,2,...,N_y$ at different
$x=x_j=jh$, $j=1,2,...,N_x,$ and $h$ is the space step. The error in scheme
(\ref{kn2}) is also second order in both space and time steps. This procedure
results in a tridiagonal set of equations (\ref{kn2}) in
$\varphi^{n+1}_{j+1,p}$, $\varphi^{n+1}_{j,p}$, and $\varphi^{n+1}_{j-1,p}$ at
time $t_{n+1}$ for each $y_p$, which are solved consistent with the boundary
conditions \cite{koo}.  Equation (\ref{gp1}) with $H=H_3$ of (\ref{h3}) is
discretized similarly:
\begin{eqnarray}\label{kn3}
\fl \frac{i(\varphi_{j,p}^{n+1}-
\varphi_{j,p}^n)}{\Delta}=-\frac{1}{2h^2}\biggr[(\varphi^
{n+1}_{j,p+1}-2\varphi^{n+1}_{j,p} +\varphi^{n+1}_{j,p-1}) 
+(\varphi^{n}_{j,p+1}-2\varphi_{j,p}^{n} +\varphi^{n}_{j,p-1})\biggr],
\end{eqnarray}
where now $\varphi^n_{j,p}$ refers to a fixed $x_j=jh$  for all $ y_p = ph$.
Using the solution obtained after $x$ iteration as  input, the discretized
tridiagonal equations (\ref{kn3}) along the $y$  direction for constant $x$ are
solved similarly. 

The iteration is started with the following normalized analytic solution
corres\-ponding to the ground states of (\ref{c}) and (\ref{d}) for spherical
and axial symmetry, respectively, with the coefficient  of the nonlinear term
set to zero
\begin{equation}
\label{ex1}
\varphi(x)={\pi}^{-0.25}2x\exp(-x^2/4)
\end{equation}
and
\begin{equation}
\varphi(x,y)=\left[\frac{\lambda}{\pi 2^{2|L|}(|L|!)^2}\right]^{0.25}
\frac{x^{1+|L|}}{2\pi}\exp[-(x^2+\lambda 
y^2)/4]  .\label{ex2}
\end{equation}
The GP equation was discretized with space step $h=0.1$ and time step $\Delta
=0.001$. For small nonlinearity the largest values of $x$ and $y$ are
$x_{\mbox{max}} = 8$, $|y|_{\mbox{max}} = 8$. However, for stronger
nonlinearity, larger values of  $x_{\mbox{max}}$ and $|y|_{\mbox{max}}$ (up to
25) are employed. The norm of the wave function is conserved after each
iteration due to the unitarity of the time evolution operator. However, it is
of advantage to reinforce numerically  the proper normalization of the  wave
function given by (\ref{n1}) or  (\ref{n2}) after each complete time iteration 
in order to improve the precision of the result.    During the iteration the
coefficient of  the nonlinear term is increased from 0 at each step by
$\Delta_1=0.0001$ for both  spherical and  axial symmetry until the final value
of  nonlinearity  ${\cal N}$  is attained at a time called time $t=0$. This
corresponds to the final solution. Then about a million time iterations of the
equation were performed which shows the stability of the result at $t=1000$.
The numerical values of the different steps and  parameters were fixed after
some experimentation in order to reduce the  numerical error. 

The maximum fluctuation of the result was found to be below few percent in the
course of a million iterations. The final wave function was found to be very
smooth at a specific time in the course of time iteration.  Consistent with the
dynamics,  small  and steady oscillation of the result appears as the nonlinear
term is increased by step  $\Delta_1=0.0001$.   This dynamical oscillation
mostly accounts for the error in the result, which can be reduced by reducing
the parameter $\Delta_1$.  We verified that there was no increase of this
numerical error with time iteration.

For large nonlinearity, the Thomas-Fermi (TF) solution of the GP equation is a
better approximation to the exact result \cite{11} than the harmonic oscillator
solutions (\ref{ex1}) and (\ref{ex2}). In that case it is advisable to use the
TF solution as the initial trial input to the GP equation with full
nonlinearity and consider time iteration of this equation without changing the
nonlinearity \cite{s3}. This time iteration is to be continued until a
converged solution is obtained.  However, in all the calculations reported in
this paper only  (\ref{ex1}) and (\ref{ex2}) are used as trial inputs.

\section{Result for Spherically Symmetric Trap}

\subsection{Stationary Problem}

First, we study the stationary problem in the spherically symmetric case. In
figure 1 (a) we plot the wave function $\phi(x,t)$ for four different values of
nonlinearity ${\cal N}=5, 100, 1000$ and 5000. The plotted wave functions are
normalized according to  $ \int_0^\infty dx\, x^2|\phi(x,t)|^2=2\sqrt 2 {\cal N}$ and
not according to (\ref{n1}). The corresponding statistical error in the wave
function is plotted in figure 1 (b). The  error was calculated through the
course of 10$^6$ time iterations with time step $\Delta =0.001$ of the final
solution generating a time interval of 1000 units of time $t$. A sample of the
wave function was chosen after 1000 iterations each. The percentage statistical
error $E$ was calculated from these samples of wave function via 
\begin{equation}\label{err} E= \frac{100}{\bar \phi
(x)}\biggr[ \frac{1}{J}\sum_{j=1}^{J} \left| \left(|\phi_j(x)| ^2
 - \bar \phi
(x)^2\right)  \right|
\biggr] ^{1/2},\end{equation} 
where $\bar \phi$ is the average  of $|\phi| $ over total number of samples
$J=1 000$. From figure 1 (b) we find that the percentage error is less than
about 0.5$\%$ for the central part of the condensate where the wave function is
sizable. It increases to about $2\%$ near the boundary where the wave function
is almost zero.  The most interesting feature of the present calculation is
that the solution remains stable even after a million iterations.   We
calculated the statistical error for the first and last 100  of the 1000
samples above. The errors so calculated were the same as those in figure 1 (b),
which shows that error does not increase with time. In other previous 
approaches it was not possible to obtain stable result after such a large
number of iterations of the GP equations \cite{s2,s4,2}.   The stability with
time iteration in the present approach will be an asset in the study of time
evolution problems which we undertake next.
\begin{figure}[!ht]
\begin{center}
\includegraphics[width=0.7\linewidth]{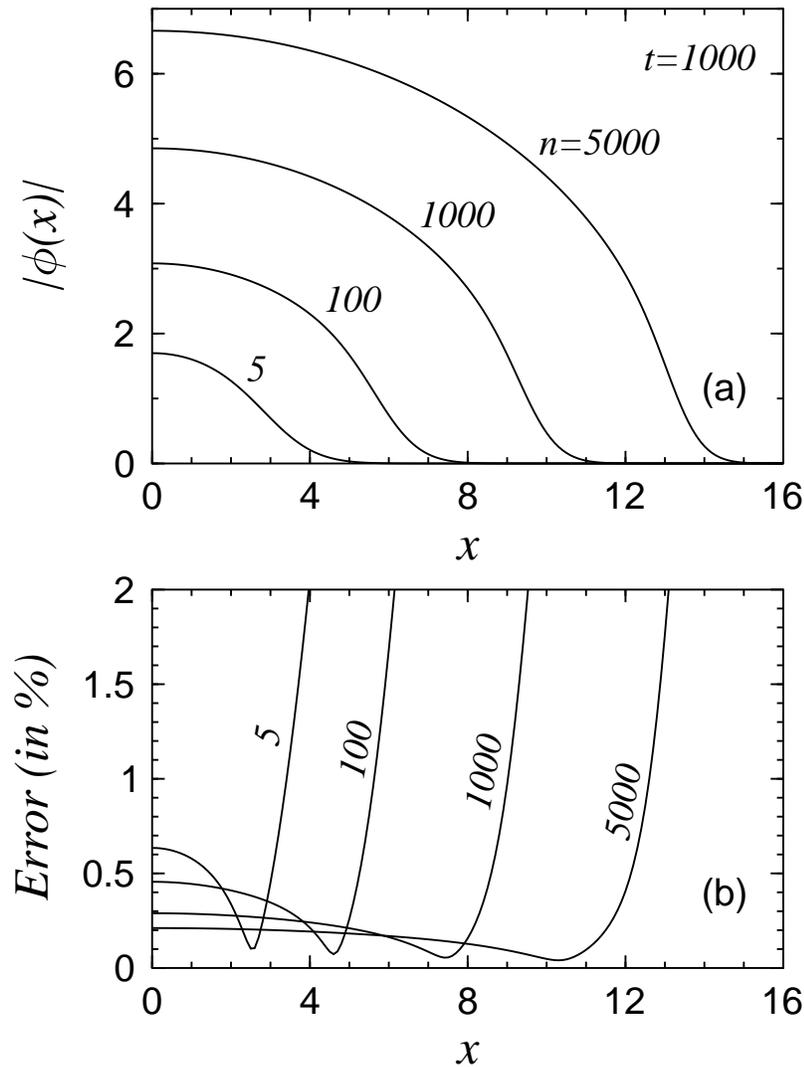}
\end{center}
\caption{The wave function in the spherically symmetric case  (a) $|\phi(x)|$
vs. $x$ at time $t=1000$ for nonlinearity ${\cal N}=5, 100, 1000, 5000$ after a
million time iterations of step $\Delta = 0.001$ of the solution at $t=0$ and
(b) the corresponding average  statistical error vs. $x$ during this interval.
The plots are labeled by the respective $n$ values and the  wave functions are
normalized as $\int_0^\infty dx x^2 |\phi(x)|^2=2\sqrt 2 {\cal N}$.}
\end{figure}

\subsection{Time Evolution Problems}

There are several interesting time evolution studies that one can pursue. After
the generation of a stable condensate one can reduce or increase suddenly the
strength of the harmonic oscillator trap by a factor and study the oscillation
of the condensate thereafter. Also, now it is possible to manipulate the
effective strength of interatomic interaction through a Feshbach resonance
after a stable condensate has been formed by changing the surrounding
electromagnetic field \cite{fb1,fb2}.  So one can increase or reduce  suddenly
the strength of the nonlinear term through a change of the scattering length in
this fashion and study the oscillation of the condensate thereafter. In both
these cases one observes the time evolution of the root mean square (rms)
radius $X$ of the condensate. This steady oscillation is best studied
theoretically by plotting   `velocity'  $\dot X$ vs.  `position' $ X$, where
dot denotes time derivative in the course of time evolution.
\begin{figure}[!ht]
\begin{center}
\includegraphics[width=0.6\linewidth]{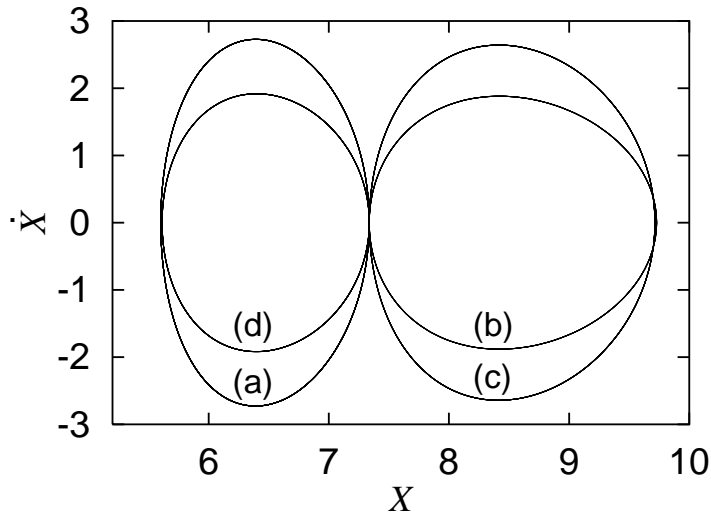}
\end{center}
\caption{Root mean square velocity $\dot X$ vs. $X$ during sustained  regular
periodic oscillation of the spherically symmetric condensate with ${\cal N}=2000$
initiated at $t=0$ by suddenly (a) increasing or (b) decreasing the harmonic
oscillator term by a factor of 2, and by suddenly (c) increasing or (b)
decreasing the nonlinear term by a factor of 2.}
\end{figure}
 
For the time evolution study we consider a previously formed condensate with
${\cal N}=2000$ prepared at $t=0$ as in the study of the stationary problem.  We
then inflict the four following changes in the system. At $t=0$ we
(a) increase or (b) decrease suddenly the coefficient of the harmonic
oscillator $x^2/4$ term in (\ref{c}) by a factor of 2. Next at $t=0$
we (c) increase or (d) decrease suddenly the coefficient of the nonlinear
term ${\cal N}$ in (\ref{c}) by a factor of 2. In all these four cases we
iterate the GP equation in time with time step $\Delta =0.001$ and observe
the system for an interval of 1000 units of time over 10$^6$ iterations.
The maximum radial distance was taken to be $x_{\mbox {max}}=25$. 
\begin{figure}[!ht]
\begin{center}
\includegraphics[width=0.5\linewidth]{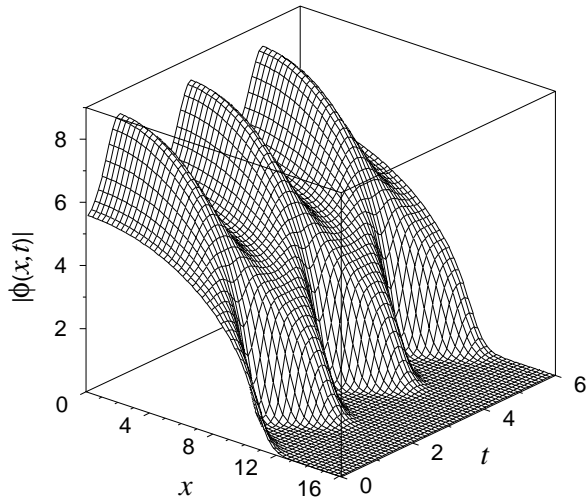}
\end{center}
\caption{The wave function $|\phi(x,t)|$ of the spherically symmetric
condensate vs. $x$ and $t$ for ${\cal N}=2000$ during breathing oscillation in
case of figure 2 (a).}
\end{figure}

In figure 2 we plot velocity $\dot X$ vs. radius $X$ in the above four
cases. We obtain a closed curve in the phase space which confirms the
periodic nature of a sustained oscillation of the condensate over 1000
units of time. The clean closed curve over such a long interval of time
assures of the low numerical error in the calculation. When the harmonic
oscillator term is doubled or the nonlinearity halved, the system is
compressed and the rms radius oscillates between its initial value and a
smaller final value.  When the harmonic oscillator term is halved or the
nonlinearity doubled, the system expands and the rms radius oscillates
between its initial value and a larger final value. The four curves of
figure 2 should originate at a central point as the initial condition on all
of them is at $t=0$, $\dot X =0$ and $X$ = constant. However, the
intersection of the pair of curves at the other extremities is accidental.  
This periodic oscillation of the wave function is clear from the phase
space diagram presented in figure 2. This oscillation is explicitly shown in
figure 3 where we plot $|\phi (x,t)|$ vs.  $x$ and $t$ in case (a) above
when the harmonic oscillator term is doubled suddenly at $t=0$ on a
previously formed condensate with ${\cal N}=2000$. The regular
breathing-mode-type pattern of oscillation of figure 3 apparently continues
for ever.  The periodic increase and decrease in central density are
accompanied, respectively, by a periodic decrease and increase in radius.  
Similar oscillation exists in the three other cases considered in figure 2
above.

We also calculated the frequency of oscillation of the rms radius $X$ in
the four cases considered in figure 2. The frequencies of simulation in
cases (a), (b), (c), and (d) above are 0.50, 0.25, 0.35, and 0.35. We are
measuring time in units of $\omega ^{-1}$ or $(2\pi \nu ) ^{-1}$, where
$\nu $ is the frequency of the harmonic trapping potential. Hence in
present units with $\omega = 1$ the trap frequency corresponding to
(\ref{c}) is $\nu =1/(2\pi)$.  We studied similar oscillation as in cases
(a) and (b) above for nonlinearity ${\cal N}=0$. In both cases the frequency of
oscillation was two times the existing harmonic oscillator frequency. As
the frequencies are increased and reduced by $\sqrt 2$ in cases (a) and
(b), two times the existing frequency for (a) is 2$\nu= \sqrt 2/\pi \simeq
0.45$ and for (b) is 2$\nu =1/(\sqrt 2 \pi )\simeq 0.23 $. In case of (c)
and (d) the frequency is unchanged and $2\nu = 1/\pi\simeq 0.32$. These
numbers compare well with the respective results of simulation, e.g.,
0.50, 0.25, and 0.35. The difference between the two sets is due to the
large nonlinearity (${\cal N}=2000$) present in simulation.

\section{Result for Axially Symmetric Trap}

\subsection{Stationary Problem}

We study the stationary problem in the axially symmetric case. All results
reported in this section are calculated with $\lambda = \sqrt 8$.   For $L=0$,
we calculated  the wave function $\phi(x,y)$ for nonlinearities up to  $ {\cal
N}=500 $ and normalized according to (\ref{5}). We also calculated the 
percentage statistical error over 100 units of time calculated from 100 samples
via (\ref{err}). For all values of nonlinearities,  the error is found to be
smaller than few percentage points almost over the entire region. As in the
spherically symmetric case, the error in the central region remained  much less
than near the peripheries.  Even for the largest  nonlinearity the convergence
is smooth and the error remained small.
\begin{figure}[!ht]
\begin{center}
\includegraphics[width=\linewidth]{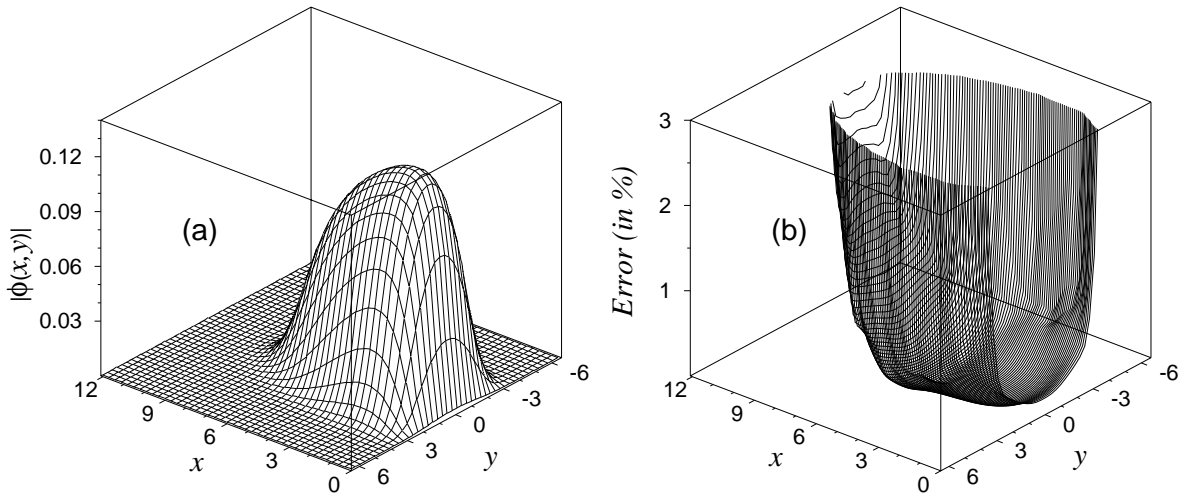}
\end{center}
\caption{The vortex wave function in the axially symmetric case (a) $|\phi(x,y)|$ vs.
$x$ and $y$ at time $t=100$ for ${\cal N}=100$ and $L=2$ after 100000 time iterations
of step $\Delta = 0.001$ after forming the solution at $t=0$ and (b) the
corresponding average  statistical error vs. $x$ and  $y$  during this
interval.  The wave function is normalized according to (\ref{5}).}
\end{figure}

We also tested our approach for vortex states, which could be  more difficult
to calculate numerically due to the $1/x^2$ term. In figure 4 (a) we plot the
vortex wave function for $L=2$ and ${\cal N}=100$. The corresponding numerical
error calculated with 100 samples over 100 units of time is plotted in figure 4
(b). Even for the vortex state $L=2,$  with a nonlinearity as large as ${\cal
N}=100$ the error is very small as we can see in figure 4.

\subsection{Time Evolution Problem}

Next we study time evolution problems with an axially symmetric trap. We
take an initial state with ${\cal N}=50$ and $\lambda = \sqrt 8$ which is
prepared as usual at time $t=0$ when we reduce suddenly the strength of
the nonlinear term by a factor of 2. This corresponds to reducing the
scattering length by the same factor, which can be realized in experiments
\cite{fb1,fb2}.  The system then starts to oscillate which may continue
for ever as in the spherically symmetric case, albeit with different
frequencies in radial and axial directions. We calculate the time
evolution of the rms values of $x$ and $y$:  $X$ and $Y$. The oscillation
is more involved and in figure 5 we plot $X$ and $Y$ vs. $t$.  Finally, we
consider again the same initial state with ${\cal N}=50$ at $t=0$ and we
double suddenly both the radial and axial harmonic oscillator trapping
terms. The system starts to oscillate after the change.  This is clearly
exhibited by plotting $X$ and $Y$ vs. $t$ in figure 6.
\begin{figure}[!ht]
\begin{center}
\includegraphics[width=0.6\linewidth]{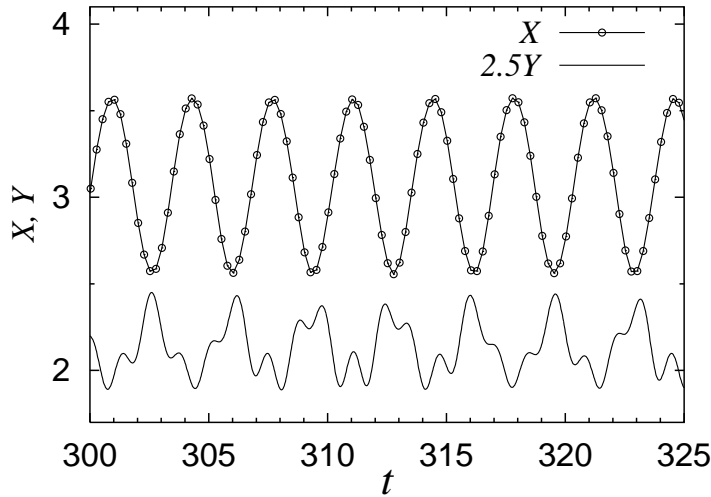}
\end{center}
\caption{Root mean square sizes $X$ and $Y$ vs. $t$ during the oscillation
of a previously formed axially symmetric condensate with ${\cal N}=50$
when the nonlinearity is suddenly reduced by a factor of 2.}
\end{figure}

We  calculated the frequency of oscillation of rms sizes $X$ and $Y$.   In case
of the simulation corresponding to figure 5 the principal frequency of
oscillation along $X$ direction is  $\nu_ 1 \simeq 0.29 $ and that in $Y$
direction is $\nu_2 \simeq 0.84$.  In case of simulation of figure 6 these
frequencies are $\nu_1 \simeq 0.42 $ and $\nu_2 \simeq 1.16$, respectively. We
also considered  the  frequency spectrum of the time variations represented in
figures 5 and 6 and we noted several  frequencies, of which the principal in
each case are, $\nu_ 1, \nu_ 2,2\nu_ 1, 2\nu_2, |\nu_ 2\pm \nu_ 1 |  $ and
$\sqrt{(\nu_ 1 ^2+\nu_ 2^2)}$. 
\begin{figure}[!ht]
\begin{center}
\includegraphics[width=0.6\linewidth]{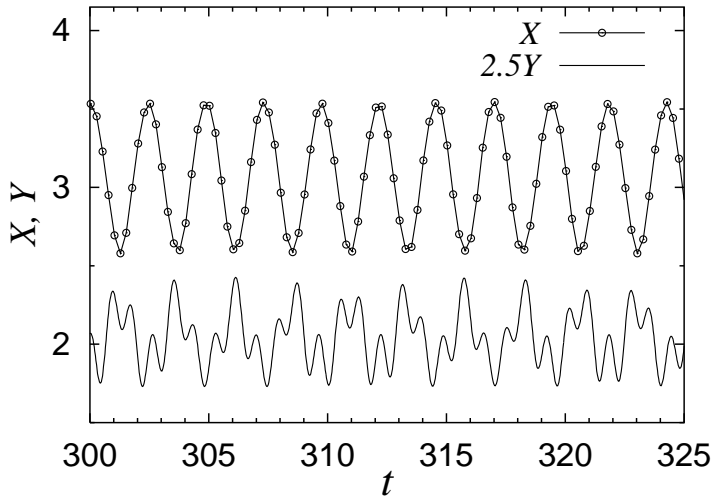}
\end{center}
\caption{ Root mean square sizes $X$ and $Y$ vs. $t$ during the
oscillation of a previously formed axially symmetric condensate with
${\cal N}=50$ when the harmonic oscillator terms are suddenly increased by
a factor of 2.}
\end{figure}

Similar oscillation has been observed in a recent experiment \cite{fb2} for a
dynamically changing axially symmetric condensate with small nonlinearity $-$
the observed frequencies being twice the existing harmonic oscillator trap
frequencies $\nu_{\mbox{radial}}$ and $\nu_{\mbox{axial}}$ along radial and
axial directions. We repeated the simulation of figure 6 with nonlinearity zero
and find that the resultant frequencies of simulation are exactly double the
existing harmonic oscillator trap frequencies $\nu_{\mbox{radial}}$ and
$\nu_{\mbox{axial}}$ along radial and axial directions. In case of figure 5
these frequencies are $2\nu_{\mbox{radial}}= 1/\pi\simeq 0.32$ and
$2\nu_{\mbox{axial}}= \sqrt 8 /\pi\simeq 0.90$ ($\nu_{\mbox{axial}} /
\nu_{\mbox{radial}} = \sqrt 8\approx 2.83$), which are to be compared with the
frequencies $\nu _1 \approx 0.29$ and $\nu _2 \approx 0.84$  ($\nu_2/\nu_1
\approx 2.90$) obtained in the present simulation, respectively. The small
disagreement between the two sets is due to the large nonlinearity of the
present simulation.  In case of figure 6 the existing trap frequencies are
increased by $\sqrt 2$ and hence the expected frequencies of oscillation are
$2\nu_{\mbox{radial}}= \sqrt 2 /\pi\approx 0.45$ and $2\nu_{\mbox{axial}}=
4/\pi \approx  1.27$, which compare well with the following results of
simulation with large nonlinearity: $\nu _1 \approx 0.42$ and $\nu _2\approx
1.16$, respectively  ($\nu_2/\nu_1 \approx 2.76$),  again the difference being
due to the strong nonlinearity. Although,  we could not relate the present
frequencies of numerical simulation  to those of the low-energy elementary
excitations of a  Bose-Einstein condensate  in an axially symmetric trap
\cite{ff}, it is interesting to note that the nonlinear  combination 
$\sqrt{(\nu_1^2+\nu_2^2)}$ appears in both the cases.

\section{Discussion}

In this paper we have presented an account of the split-step Crank-Nicholson
method for the solution of the GP equation, where the nonlinear and all
nonderivative terms are treated in a separate step essentially exactly via
(\ref{al1}). In previous applications \cite{s21,s2,s31,s32,s3,s4} of the
Crank-Nicholson method to the GP equation these terms were included in the
finite-difference scheme for the space derivatives. In the pioneering
application of this approach Ruprecht et al \cite{s21} noted that  the presence
of the nonlinear potential sets a limit to the numerical stability of the
method and limits the  ability to find the wave function for large
nonlinearity. For the spherically symmetric case, Ruprecht et al \cite{s21}
produced results for values of nonlinearity ${\cal N} \equiv Na/l$ upto
$25\sqrt 2$. For the axially symmetric case, the maximum of $\cal N$ employed
by  Dalfovo and Modugno \cite{s32} and by Holland and Cooper \cite{s3} were 20
and 5, respectively. Using a different method Chiofalo et al \cite{3y} produced
results for values of $\cal N$ upto 6. However, unlike in these other
methods,   in the present method  the results remain stable for large  $\cal
N$,

At this point we must comment on the potentially powerful split-operator method
with a pseudo-spectral scheme for derivatives \cite{ps1}, often used in solving
nonlinear Schr\"odinger-type equations. In the pseudo-spectral scheme the
unknown function is expanded accurately in terms of some known polynomial (e.
g. Chebyshev polynomial) over a set of grid points. As the number of terms in
this expansion increases, the discrete representation of the function and its
derivatives becomes increasingly accurate.  In the present Crank-Nicholson
method the space derivative is approximated by a simple three-point
finite-difference scheme.  To the best of our knowledge so far there has been
no systematic application of the pseudo-spectral method to the solution of the
GP equation, specially for the axially symmetric case \cite{ps2}. Whether the
added numerical effort in the pseudo-spectral method is compensated by the
increased accuracy can only be found after a systematic comparison of the two
approaches. This remains a problem of future interest.

\section{Summary}

In this paper we propose and implement a split-step method for the
numerical solution of the time-dependent nonlinear GP equation under the
action of a trap with both spherical and cylindrical symmetries by time
propagation starting with the initial input solution for the harmonic
oscillator problem. The full Hamiltonian is split into the derivative and
nonderivative parts. In this fashion the time propagation with the
nonderivative parts can be treated very accurately. The spatial derivative
parts are treated by the Crank-Nicholson method. In the axially symmetric
case the spatial derivatives in the radial and axial directions are dealt
with in two independent steps. This split-step method leads to highly
stable and accurate results. The final result remains stable for millions
of time iteration of the GP equation.

We applied the above method for the numerical study of certain stationary and
time-evolution problems with spherical  and axial traps.  In the spherical case
stationary solutions with nonlinearity  ${\cal N}\equiv Na/l=5,100,1000,$ and
5000 showed accurate result with small numerical error. For a condensate with  
${\cal N}=2000$ we  studied certain time-evolution problems. We studied the
resulting oscillation of the condensate when at $t=0$ the trap frequency or the
scattering length is altered suddenly. In all cases a clean periodic motion of
the root mean square radius was noted.

In the axially symmetric case the  stationary solutions for $L=0$ with
nonlinearity ${\cal N}\equiv Na/l=50 $  and 300 showed accurate result. A
calculation of the vortex state for $L=2$  with nonlinearity  ${\cal N}=100$
showed equally precise and stable results. The numerical error was less than
one percent in the central region and a couple of percents in the peripheral
region. We also performed two time-evolution studies for ${\cal N}=50$ when the
harmonic oscillator trap is doubled or the scattering length is reduced
suddenly.  The root mean square sizes along radial and axial directions showed
clean periodic oscillation  which is studied in detail. The present method
seems to  be very attractive for studying various dynamical evolution problems
with dissipative (imaginary) interaction, which can not be handled efficiently 
by other methods such as variational and Thomas-Fermi methods. We have already
made such an application in the study of chaotic dynamics of the Bose-Einstein
condensate  using the Gross-Pitaevskii equation with dissipative interaction
\cite{fc2}.

\ack

The work is supported in part by the Conselho Nacional de Desenvolvimento
Cient\'\i fico e Tecnol\'ogico and Funda\c c\~ao de Amparo \`a Pesquisa do
Estado de S\~ao Paulo of Brazil.

\end{document}